\documentclass[a4papper]{article}

\usepackage[dvips]{color}
\usepackage{graphicx}
\usepackage{subfigure}
\usepackage{graphics}
\usepackage{rotating}
\usepackage[cp850]{inputenc}
\usepackage[spanish,english]{babel}
\usepackage{amsfonts,amsbsy,amsmath}
\usepackage{epsfig}
\usepackage{multicol}
\usepackage{colortbl}
\usepackage{pstricks,pst-grad}

\usepackage{bm}% bold math
\usepackage{amssymb}
\usepackage{xspace}
\usepackage{hyperref}
\usepackage{amsmath}
\usepackage{ulem}

\input epsf

\newcommand{\pp}{\ensuremath{\rm p\!+\!p}\xspace}

\newcommand{\ppb}{\ensuremath{\rm p\!+\!Pb}\xspace}

\newcommand{\pt}{\ensuremath{p_{\rm{T}}}\xspace}

\newcommand{\pbpb}{\ensuremath{\rm Pb\!+\!Pb}\xspace}

\newcommand{\xitopi}{\ensuremath{{\rm (\Xi^{-}+\Xi^{+})}/(\rm \pi^{-}+\pi^{+})}\xspace}
\newcommand{\otopi}{\ensuremath{{\rm (\Omega^{-}+\Omega^{+})}/(\rm \pi^{-}+\pi^{+})}\xspace}

\newcommand{\ltopi}{\ensuremath{{\rm 2\Lambda}/(\rm  \pi^{-}+ \pi^{+}  )}\xspace}

\begin{document}

%\begin{frontmatter}
%\frontmatter
\title{\bf{ The energy density representation of the strangeness enhancement from \pp to \pbpb}}

%\tnotetext[mytitlenote]{ecuautle@nucleares.unam.mx, guypaic@nucleares.unam.mx}

\author{E. Cuautle and G. Pai\'c \\  Instituto de  Ciencias
  Nucleares, Universidad Nacional  Aut\'onoma de M\'exico, \\ Apartado
  Postal 70-543, Ciudad de M\'exico  04510}  \maketitle{}
%\address{Instituto de Ciencias Nucleares, Universidad Nacional   Aut\'onoma de M\'exico}


\begin{abstract}
The energy density is the prime parameter to define the deconfinement
of quarks and gluons occurring in collisions of heavy ions. Recently,
there is mounting evidence that many observables in proton-proton
collisions behave in a manner very similar to the one observed in
heavy ions. We present as an additional piece of evidence,  a scaling
of the strange particle yields as a function of the energy density of
the  three collision systems: \pp \ppb and \pbpb,  using the latest
results of the ALICE collaboration. 
\end{abstract}

%\begin{keyword}
%\texttt{p+p, p-pb and Pb-Pb collisions,  \sep Energy density \sep %strangeness}
%\end{keyword}

% \end{frontmatter}

%\linenumbers

\section{Introduction}
Since the very beginning of the searches for quark gluon plasma,  the
production of strange and multi-strange baryons has been in
the focus of the experimental community,  the reason being that one
expected to observe an enhancement of the strangeness production in
heavy ion collisions compared with the production in  collisions of protons or
lighter systems\cite{Rafelski82,Rafelski82Erratum,koch}.
On the other hand, it is generally accepted that strange particle
densities and their ratios can be, for heavy ion collisions at full AGS energy and higher, well described
 by the thermal model\cite{thermalmodel1,thermalmodel2} in a grand-canonical ensemble while  in small colliding  systems 
the canonical description is needed\cite{Braun-Munzinger}.
Recently the ALICE Collaboration has %published\cite{aliceplb465-1999}
published\cite{alice-2016,alice-plb758-389-2016}
 two interesting
articles where the production of strangeness with respect to the pion
production is reported for several strange and multi-strange particles
for three colliding systems: \pp, \ppb and \pbpb. They report the relative enhancements in function of the multiplicity.
\noindent
The papers give an important message that the integrated yields of strange and
multi-strange particles relative to pions 
increase significantly with multiplicity and the results cannot be reproduced by any of the models commonly used.

\noindent
The present situation as well as the mounting evidence for collective
effect has prompted us to examine the possibility of a scaling of the
strangeness production with the energy densities achieved in the various colliding 
systems. Observing such scaling, would raise many interesting questions about the underlying
processes. It might be  noted that the scaling idea was presented\cite{Wang2002},  using a variable which is related to initial energy to describe the kaon to pion ratio from different colliding systems and  different energies.
\noindent
Recently, a similar approach was shown\cite{Castorina} for minimum bias values,
where they conclude that the degree of strangeness suppression in hadronic and nuclear
collisions is fully determined by the initial energy density.  In Ref.\cite{paiccuautle2016}, we have already presented the results for the ratio of cascades to pion production.

\noindent
In the present work we  study the relative enhancements of strangeness production for each multiplicity bin,  using the Bjorken energy density approach\cite{Bjorken83} {\it i. e.} 
the energy density reached in colliding systems, assuming that a
thermal equilibrium is achieved.
\noindent

The work is organized as follows. In the first part we explain the
way we calculate the energy density of the systems in collision. In
the second part we apply the scaling to the ALICE results for \pp, \ppb
and finally \pbpb collisions for the strange and multi strange
baryons. In the third  part we present the results and the discussion. Finally  some conclusions are drawn.

\section{Energy density}
The initial Bjorken energy density approach provide a relation of the
transverse energy and the covering area of the thermalized system produced in  hadron/nucleus collisions: 
\begin{eqnarray}
\epsilon \sim \frac{dE_{T}/d y }{\pi R^{2} \tau }
\end{eqnarray}

%where the $\tau$ is the hadronization time. 
\noindent
where the transverse energy function of the rapidity
can be expressed in terms of the pseudo-rapidity  and  the average
transverse momentum distribution, $dE_{T}/d y  \sim \;\; < \pt> dN/d\eta$. The hadronization time, $\tau$,  is essentially a rather unknown factor so that we actually compute the value
%, so for our case we have
\begin{eqnarray}
\epsilon \tau & = &  \frac{3}{2}\frac{<\pt^{k} > dN^{k} /d\eta }{A^{k} },
\label{eq:energy}
\end{eqnarray}

\noindent
where $k$ represent the colliding systems \pp, \ppb or \pbpb. In this way $\epsilon \tau$  is computed for  each multiplicity bins where  the  ratios strangeness to pions are measured,  $A^{k}$ is the area of the thermalized system mentioned above, and  $dN^{k} /d\eta $ is the charged particle density.

Our calculation were done using each multiplicity bins where strange and multi-strange ratio was published, from \pp\cite{alice-2016}, \ppb\cite{alice-plb728-25-2014} and \pbpb\cite{alice-plb728-216-2014, plb728-erratum}, at 7, 5.02 and 2.76 TeV respectively, and with $ \mid \eta \mid \, < 0.5$. The average transverse momentum, $< \pt> ^{k}$, is for for all charged hadrons, different  for each colliding system. The values to compute Eq.\ref{eq:energy} were taken from ALICE measurements\cite{ALICE-plb727-371-2013}, which correspond to the same energy as data of the multi-strange ratios, but in pseudorapidity range of $ \mid \eta \mid \, < 0.3$.

Up to here, one of the main question are the values of the radii used to calculated the thermalized systems created.
This area can be calculated by several approaches, the  Glauber model for instance, which allow the to study possible fluctuations of this area, however, 
we decide to use experimental values of the invariant radii\footnote{We are aware that the radii used here refer to the kinetic freezout and are therefore somewhat larger than the proper thermalization ones. However we believe that a coherent approach for all systems does no affect the result as long as one would not seek to determine the real energy density value} obtained from momentum correlations in the experiment\cite{alice-radii}.
This reference shows radii for the pions source from two and three cumulant methods, with small differences  between the results for \pp and \ppb (of the order of 10-15 \%.) so that we decided to use the same radii extracted for \pp for the two systems using the functional form given in Eq.\ref{eq:R_pp}  for \pp and \ppb, and Eq.\ref{eq:R_pbpb} for \pbpb:

\begin{equation}
R_{\pp}  \; = \; 0.405 \frac{dN_i}{d\eta}^{1/3} + 0.332 
\label{eq:R_pp}
\end{equation}
%\begin{equation}
%R_{\ppb} \; = \; 0.585 \frac{dN_i}{d\eta}^{1/3} + 0.540\
%\label{eq:R_ppb}
%\end{equation}
\begin{equation}
R_{\pbpb} \; = \; 0.772 \frac{dN_i}{d\eta}^{1/3} + 0.049
\label{eq:R_pbpb}
\end{equation}

\noindent
where the index  $i$ refers to  multiplicity bins. 
We would like to add that we have tried several other methods to calculate the energy density and the functional form does not substantially influence the conclusions presented here.

\section{Results}
\begin{figure}[htbp]%[t] %[htbp]
\begin{center}
      \includegraphics[width=0.80\textwidth]{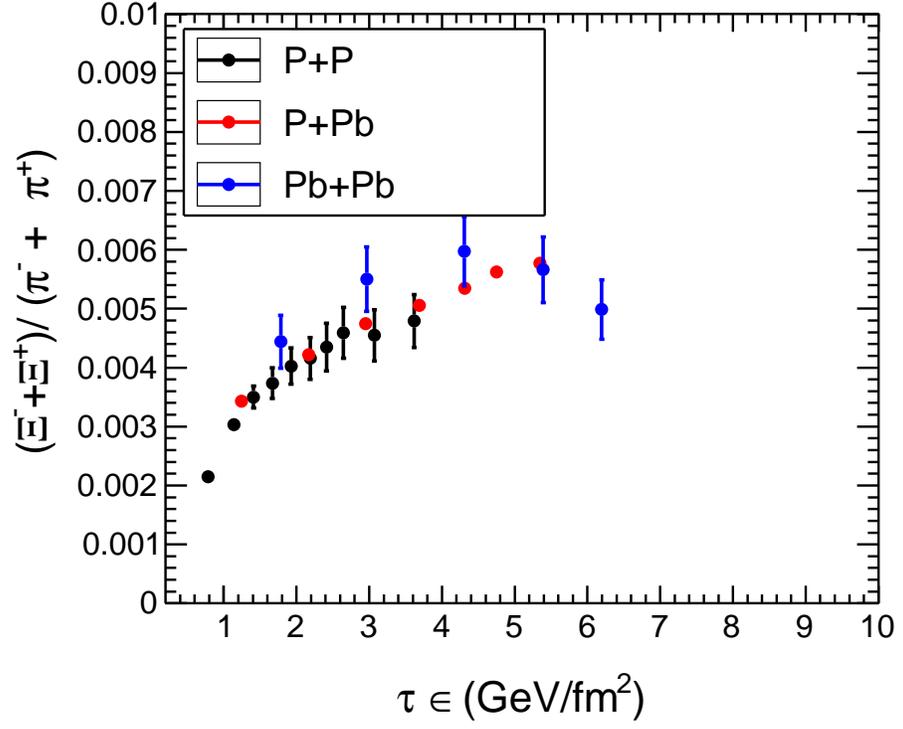}
   \caption{(Color online) \xitopi ratios at 7 (\pp), 5.02 (\ppb) and 2.76 (\pbpb) TeV as function of energy density.}
  \label{Fig1}
\end{center}
\end{figure}

\begin{figure}[htbp]%[t] %[htbp]
\begin{center}
    \includegraphics[width=0.80\textwidth]{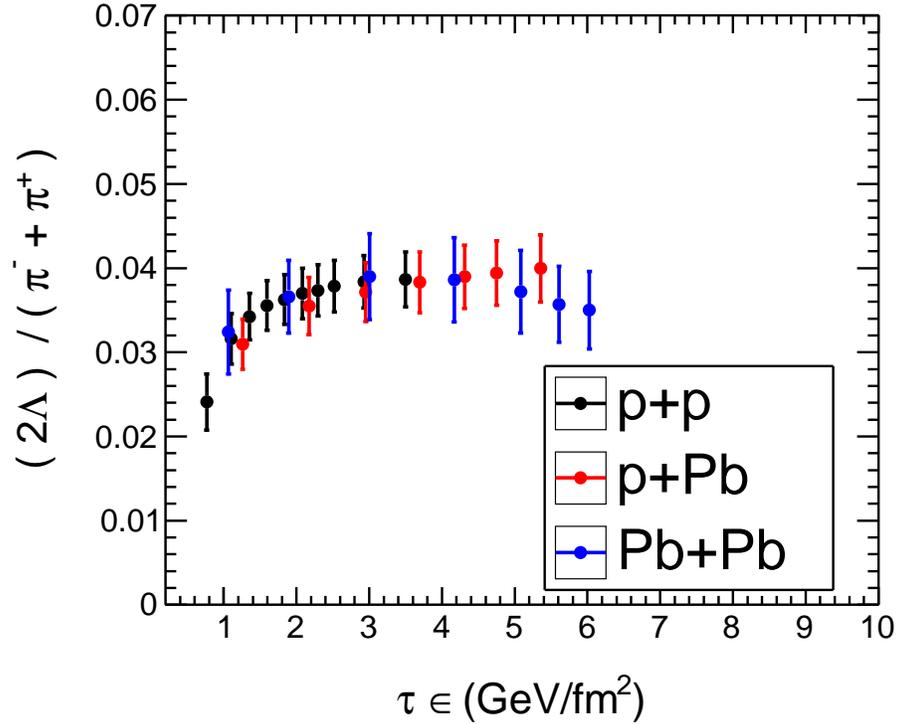}
   \caption{(Color online) \ltopi ratios at 7 (\pp), 5.02 (\ppb) and 2.76 (\pbpb) TeV  as function of energy density.}     
  \label{Fig2}
\end{center}
\end{figure}

\begin{figure}[htbp]%[t] %[htbp]
\begin{center}
    \includegraphics[width=0.80\textwidth]{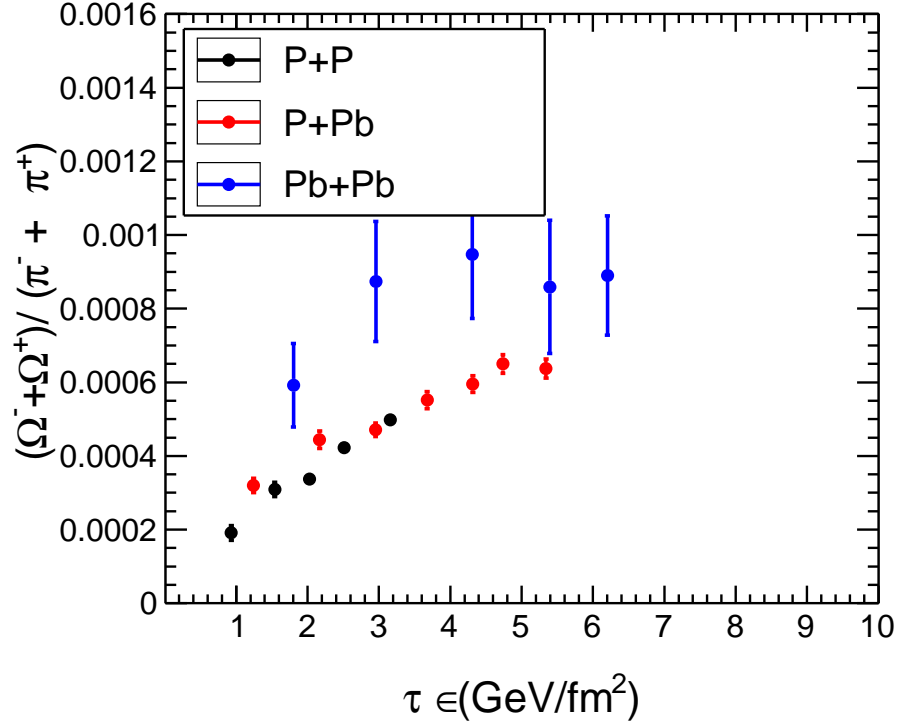}
   \caption{(Color online) \otopi ratios at 7 (\pp), 5.02 (\ppb) and 2.76 (\pbpb) TeV  as function of energy density.}
  \label{Fig3}
\end{center}
\end{figure}

In order to estimate the ratios: \xitopi, \ltopi and \otopi  versus initial energy density, we use the Eqs.\ref{eq:energy} and the radii extracted from results of \pp, Eq.\ref{eq:R_pp} for the results from \pp and \ppb, while for those from \pbpb the Eq.\ref{eq:energy} and Eq.\ref{eq:R_pbpb} are used. The  Figs.\ref{Fig1}, \ref{Fig2} and \ref{Fig3} show 
these ratios for the systems studied.
We can observe in the three cases the same trend of the enhancements, a nearly perfect scaling  among the three different systems.\\

\noindent
\noindent
The \otopi ratio seems to indicate more production of strangeness in \pbpb systems, but it is not completely conclusive due to  the large errors.

\section{Conclusion}
The energy density plotted using the  Bjorken formula show that at the same energy $\epsilon \tau$, one obtains the same enhancement regardless of the system measured.
The scaling of the relative production of strange and multi strange baryons looks reasonably successful for all the baryons presented except for the case of the Omega baryons where the large errors prevent to draw a final judgement. However it seems that the production in \pbpb collisions is more copious. \\
In our opinion this has important consequences, for instance,
it is comforting the idea that, even in small systems, the thermalized energy density may be a relevant parameter to describe the production of strangeness. 
The results are in agreement with those from the
holographic duality studies that demonstrate the applicability of hydrodynamics to small systems at sufficiently high initial temperatures\cite{chesler}. On the other hand, the results presented do not indicate a clear tendency towards a limiting temperature as predicted by the thermal models.

\section*{Acknowledgments}
The authors acknowledges useful discussions with Antonio Ortiz.
Support for this work has been received from CONACYT under the grant
No. 270067 and  from DGAPA-UNAM under PAPIIT grants  IN108414.

%\section{Bibliography styles}
\bibliographystyle{unsrt}
%\section*{References}

%\begin{thebibliography}{99}
\bibliography{references}
%\end{thebibliography}

%\include{references}
\end{document}